\documentclass[journal=nalefd,manuscript=letter]{achemso}

\author{A.~A.~Kaverzin}
\email{ak328@exeter.ac.uk}
\author{S.~M.~Strawbridge}
\author{A.~S.~Price}
\author{F.~Withers}
\author{A.~K.~Savchenko}
\author{D.~W.~Horsell}
\affiliation{School of Physics, University of Exeter, Exeter, EX4 4QL, UK}

\title{Electrochemical doping of graphene}

\begin{document}

\begin{abstract}
The electrical properties of graphene are known to be modified by chemical species that interact with it. We investigate the effect of doping of graphene-based devices by toluene (C$_{6}$H$_{5}$CH$_{3}$). We show that this effect has a complicated character. Toluene is seen to act as a donor, transferring electrons to the graphene. However, the degree of doping is seen to depend on the magnitude and polarity of an electric field applied between the graphene and a nearby electrode. This can be understood in terms of an electrochemical reaction mediated by the graphene crystal.
\end{abstract}


Graphene is a single atomic layer of the crystal graphite. It is a semiconductor with a zero energy band-gap and a linear energy spectrum. As a result, its electrical properties are highly unusual \cite{CastroNetoRMP81}. Although graphene-based transistors are predicted to demonstrate the highest mobility of charge carriers at room temperature, $\sim10^5$\,cm$^2$/Vs \cite{HwangPRB77}, this has yet to be realised in experiments as the electrical properties are strongly modified in the presence of other materials. These can act either as acceptors or donors when they come into contact with the graphene surface thereby changing its charge carrier density. The details of the interactions of different molecules with graphene are not well understood and yet are of major importance for practical device applications \cite{GeimScience324}.
	
The effects of several inorganic\cite{WehlingNL8,SchedinNatMat6,PiPRB80,HummelPSSB247} and organic\cite{OhnoNL9,DasChemComm2008,DanNL9,DongSmall5} molecules on the electrical conduction of graphene have been demonstrated. They were shown to act as dopants, the change in carrier density depending on the type and concentration of the chemical species. Moreover, the conductance of a graphene-based transistor appears to be sensitive to the presence of individual molecules of NO$_{2}$ \cite{SchedinNatMat6}. It has been predicted, though until now not experimentally demonstrated, that organic molecules can cause not only this simple type of `molecular' doping but also doping as a consequence of electrochemical reactions\cite{SahaPRB80,Pinto10030624}. Carbon nanotube-based devices have been shown to be effective sensors for such reactions (for a recent review, see \cite{HuJSensors2009}) so we would expect graphene to hold even greater promise for practical device applications.

\begin{figure}
\includegraphics[width=.6\textwidth]{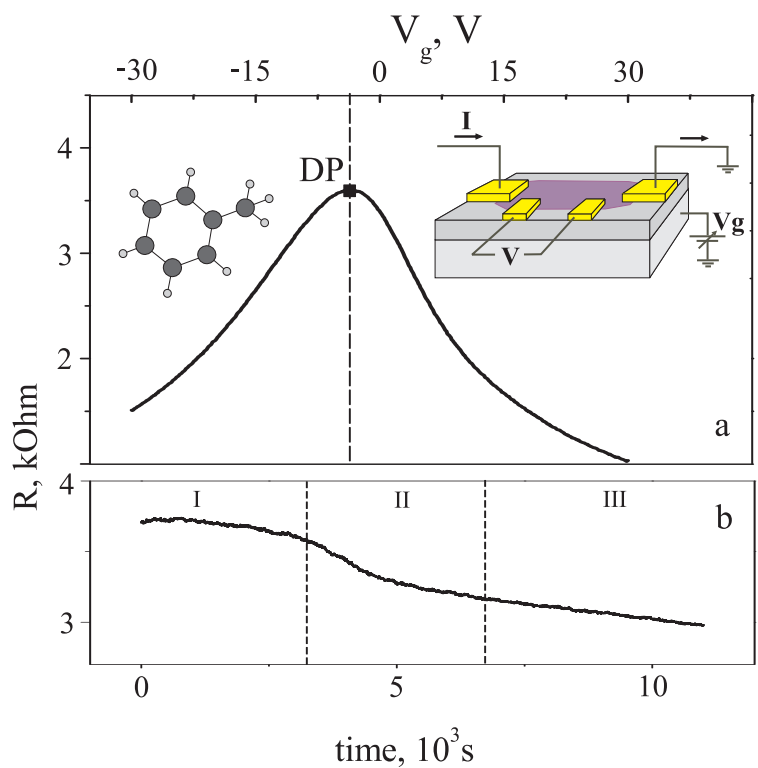}
\caption {(a) Resistance $R$ of a typical graphene device as function of the gate voltage $V_g$. The position of the Dirac point (DP) is indicated. Left inset: structure of the toluene molecule. Right inset: schematic of the device and circuit. (b) Change of the resistance as a function of time after adding toluene vapour with no applied gate voltage. Three distinct time intervals are highlighted.}
\label{fig:one}
\end{figure}

In this work we report the effect of doping graphene with the aromatic molecule toluene, C$_{6}$H$_{5}$CH$_{3}$, \ref{fig:one}(a, inset). We show that the way in which this doping occurs is significantly different from the simple molecular doping of graphene studied earlier (where transfer of electrons occurs directly between graphene and the HOMO or LUMO energy levels of the dopant). The observation of hysteresis and enhancement of the doping effect by an electric field produced by a nearby gate electrode suggest that an electrochemical reaction lies at the origin of the doping process. For our experimental conditions, we determine the energy scale of the reaction responsible for the graphene doping by toluene. Our results indicate that the dipole moment in toluene plays a role in the origin of this effect.


Graphene flakes were produced by micromechanical cleavage \cite{NovoselovScience306} of natural graphite and deposited on a degenerately doped silicon substrate covered by 300\,nm silica. The flakes were confirmed to be single layer by Raman spectroscopy \cite{FerrariPRL97}. Electrical contacts (Cr/Au) were then made to each flake. The carrier density $n$ was tuned by applying a voltage $V_{g}$ between the graphene flake and the conducting silicon substrate which acts as the gate. (The density $n$ is determined by the gate--flake capacitance: $n$\,(cm$^{-2}$)$=7.2\cdot 10^{10} V_{g}$\,(V).) The right inset to \ref{fig:one}(a) shows a schematic of a typical sample. A total of seven samples were studied in detail, their dimensions ranging from 2 to 18\,$\mu$m in width and length.

Measurements were performed at room temperature in a sealed chamber connected to a vacuum pump and a pure helium gas source. Before exposing the sample to toluene the device was annealed in vacuum at 140$^{\circ}$\,C for one hour to remove (as far as possible) contaminants from the surface. Exposure to toluene was then performed in an inert helium atmosphere. The source of toluene vapour was from the natural evaporation from a liquid reservoir placed within the chamber immediately under the sample. Under these conditions the areal coverage of toluene on the graphene surface was estimated to be small (less than $1\%$), however, exact knowledge or control of the coverage was not important in this case for the observed effect or analysis.

\ref{fig:one}(a) shows the resistance as a function of the gate voltage in the absence of doping. The half filled outer shell of electrons in carbon leads to the Fermi level in pure graphene lying at the (gapless) point between the conduction and valence bands---the Dirac point. At this point the net density of electron states is zero and the resistance of a graphene sample is maximal. For an undoped sample this occurs at $V_g=0$. When a negative (positive) $V_g$ is applied, the Fermi level is shifted down (up) and the sample resistance decreases due to adding holes (electrons) to the channel. In a doped sample, the resistance peak is shifted from $V_g=0$ as electrons or holes are added to the channel by the dopant.

\ref{fig:one}(b) shows the effect of toluene on the resistance of a graphene device at $V_g=0$ as a function of time. The addition of toluene into the chamber is seen to change the resistance over a timescale of hours. There are three distinct intervals. First, there is an initial `delay' of $\sim10^3$\,s between adding toluene and the most significant change of the resistance. We ascribe this to the time taken for toluene to reach and form a layer at the graphene surface. This delay is only observed when the (annealed) sample is first exposed to toluene in the chamber. Second, there is an interval of time where the change in $R$ is large. Here the toluene is having its greatest effect and will be discussed in detail below. Third, beyond $7\cdot 10^3$\,s a drift in the value of $R$ is observed. This can be attributed to a gradual increase in the areal coverage of toluene on the surface.

\begin{figure}
\includegraphics[width=.6\textwidth]{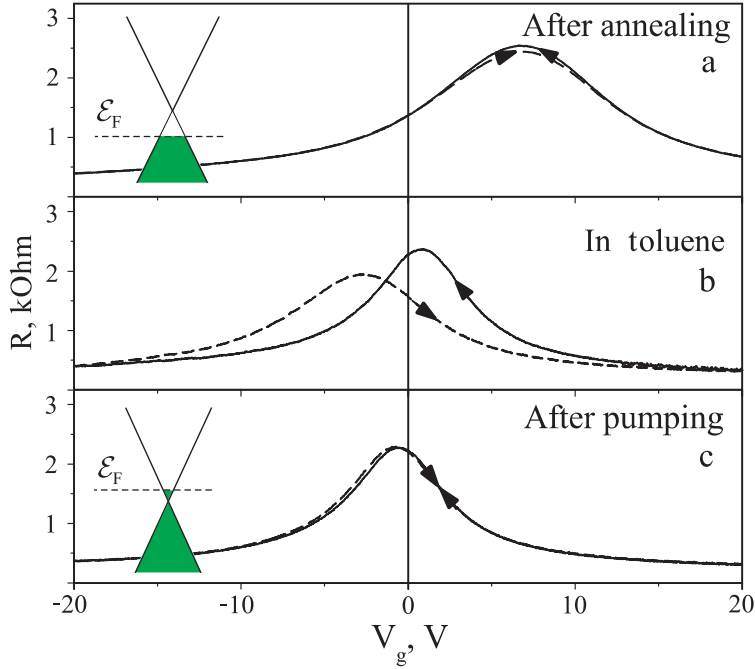}
\caption {$R(V_{g})$ at different stages of the doping experiment: (a) after annealing the sample in vacuum; (b) after doping by toluene; (c) after pumping out toluene vapour. Insets show the linear energy dispersion curve for graphene where the occupation of states is indicated at $V_g=0$.}
\label{fig:two}
\end{figure} 

\ref{fig:two}(a) shows that before adding toluene the annealed sample is doped with holes. The addition of toluene shifts the resistance peak towards negative values of $V_g$ indicating that it acts as a donor, \ref{fig:two}(b). However, the effect of toluene is not simply to shift the peak. In addition, hysteresis is observed in the $R(V_g)$ curves: the exact position of the resistance peak depends upon the direction of the $V_g$ sweep. The hysteresis is not a transient effect and does not disappear when the sweeping rate is decreased. (Hysteresis resulting from a simple time lag in the system would cause the reverse of the two curves in \ref{fig:two}(b).) When the toluene is pumped out of the chamber the doping effect remains but the hysteresis disappears, \ref{fig:two}(c). (We found experimentally that the doping effect can only be removed when the sample is heated above $\sim200^{\circ}$\,C.)


Let us first consider the doping effect of toluene. Calculations have shown \cite{Pinto10030624} that the Fermi level of graphene with a toluene molecule on its surface is not shifted with respect to the Fermi level of pristine graphene. Therefore, the doping mechanism has to either involve other chemical species or be a more complicated process than simple molecular doping. It is known that chemical residues originating from the device fabrication process exist on the flake. Some of these cannot be removed by annealing, and any that act as dopants will cause an offset of the resistance peak from $V_g=0$. Such an offset is seen in \ref{fig:two}(a). The peak here is shifted to the right, which indicates initial doping by holes. There are several chemical species that could give rise to such doping: gold and chromium atoms from the evaporation of the contacts, PMMA residues from the lithographic processing \cite{DanNL9}, and water trapped between the graphene and silica surface \cite{JoshiJPCM22}.

Gold and chromium are unlikely to cause doping in our samples. To do so they must occur as individual atoms on the graphene surface \cite{Pinto10030624}, which is highly unlikely to result from evaporation. To understand the importance of PMMA experimentally, we fabricated samples that involved no PMMA in their processing. For these samples, the Au/Cr contacts were evaporated through a shadow mask formed of a thin copper membrane containing two 200\,$\mu$m wide holes spaced 18\,$\mu$m apart. The sample was not immersed in any solvents during preparation so that there was no chance for PMMA or any other residues to contaminate the flake (though this does not exclude possible atmospheric contaminants). The effect of toluene doping of samples created by this shadow-mask technology has not shown any qualitative difference to that seen for lithographically processed samples, though the initial doping of the sample is in general lower. A surface layer of water, therefore, is the most likely origin of the hole type doping, and, in addition, can be a factor in the mechanism of doping by toluene (discussed below). 

\begin{figure}
\includegraphics[width=.6\textwidth]{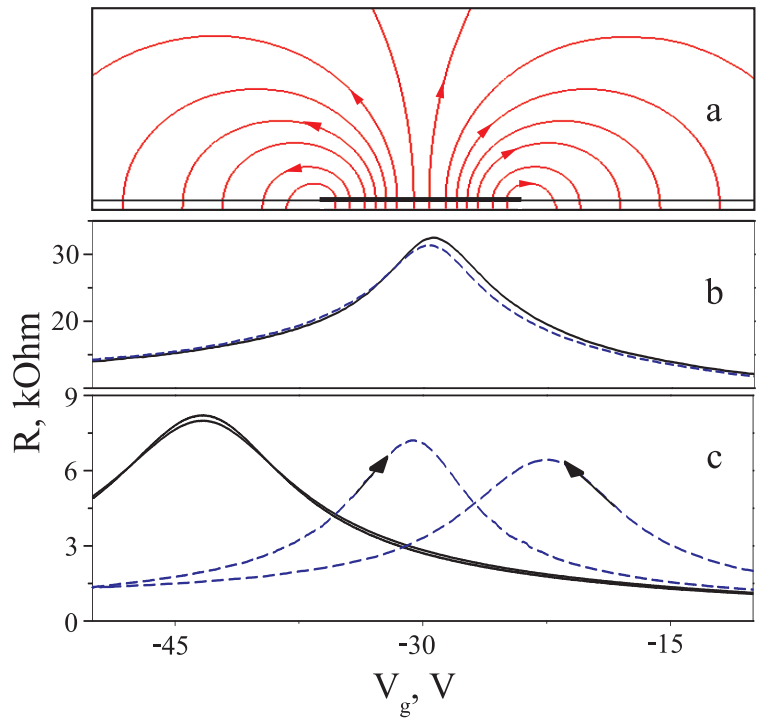}
\caption {(a) Electric field calculated for our device geometry with an applied voltage between the graphene flake (shown as a thick black line) and doped silicon gate (bottom of image). (b) $R(V_g)$ measured in the absence (solid line) and presence (dashed line) of naphthalene. (c) $R(V_g)$ in the absence (solid line) and presence (dashed line) of water.}
\label{fig:water}
\end{figure}

Let us now consider the hysteretic behaviour. While doping by toluene occurs at $V_g=0$, \ref{fig:one}(b), applying a gate voltage to the system strongly affects the degree of doping. Toluene is a dipolar molecule, with a dipole moment $p=1.2\times 10^{-30}$\,C$\cdot$m, and as such will be sensitive to an applied electric field. The gate voltage creates an electric field not only uniformly below but also nonuniformly above the graphene flake, \ref{fig:water}(a). The `stray' field extends into the volume above the flake with a strength around $10\%$ (within $\sim$100\,nm) that of the uniform field below. Such a field is sufficient to influence the reactivity of toluene \cite{Schmickler}. To experimentally test whether the dipolar nature of the toluene is significant, we repeated the experiments with other molecules. Naphthalene, C$_{10}$H$_{8}$, is a symmetric molecule consisting of two fused benzene rings and has zero dipole moment. When naphthalene was introduced into the chamber we observed no doping effect, \ref{fig:water}(b). This result suggests that the origin of the effect is not due to a $\pi$--$\pi$ stacking interaction \cite{GrimmeACIE2008}. Water, which is dipolar, $p=6.2\times 10^{-30}$\,C$\cdot$m, was also investigated. In the presence of water vapour, doping and hysteretic behaviour was observed, \ref{fig:water}(c) (a feature also seen in the results of \cite{LafkiotiNL10,LohmannNL9}). As for toluene, pumping out the water vapour from the sample chamber eliminated the hysteresis. Further experiments with another dipolar molecule, aniline, indicate the presence of this effect, though this was not studied in detail. All these experiments indicate that the hysteresis and doping effects result from the same mechanism, and that this mechanism occurs more readily in doping with dipolar molecules. Therefore, investigation of the hysteresis can be used to explore the doping mechanism in more detail.

\begin{figure}
\includegraphics[width=.6\textwidth]{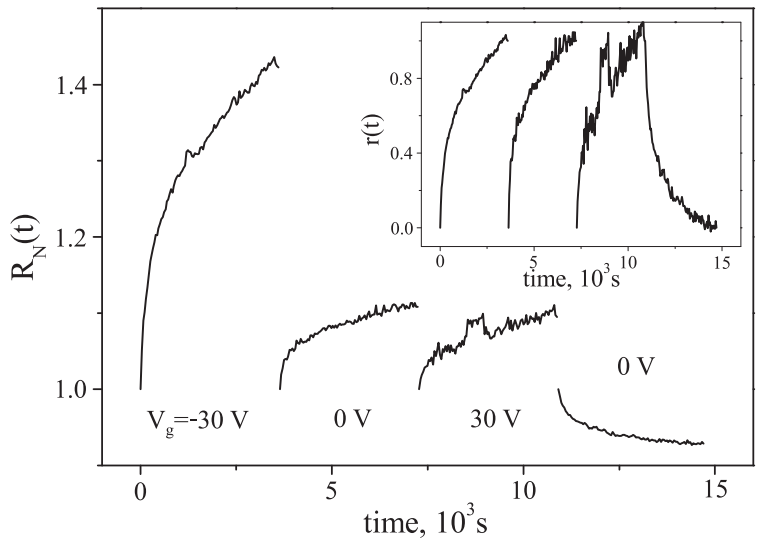}
\caption {Time dependence of the (normalised) resistance, $R_N$, after rapid changes in the gate voltage: from $V_g=0$ to $-30$\,V, to 0, to $+30$\,V, and back to 0. Inset: data as in main figure but normalised by the full resistance range, $r=(R-R_\mathrm{min})/(R_\mathrm{max}-R_\mathrm{min})$.}
\label{fig:four}
\end{figure}

\ref{fig:four} shows the resistance (normalised by its initial value for each curve) as a function of time in the presence of toluene vapour. First, the sample was stabilised for three hours at $V_{g}=0$. Then the gate voltage was quickly swept to $-30$\,V and the time dependence measured. Surprisingly, after changing $V_g$, the resistance of the sample changes significantly from its new initial value. The same is true upon rapidly sweeping to other gate voltages, shown as a sequence in the figure. This indicates that the doping depends on the applied gate voltage: for simple molecular doping no change would be expected, and in the case of a simple time lag the evolution of the resistance would be in the opposite direction for all but the second curve in the figure. It follows that the number of electrons transferred from toluene to graphene depends on $V_g$, and thus so do the Fermi level and position of the Dirac point.

\begin{figure}
\includegraphics[width=.6\textwidth]{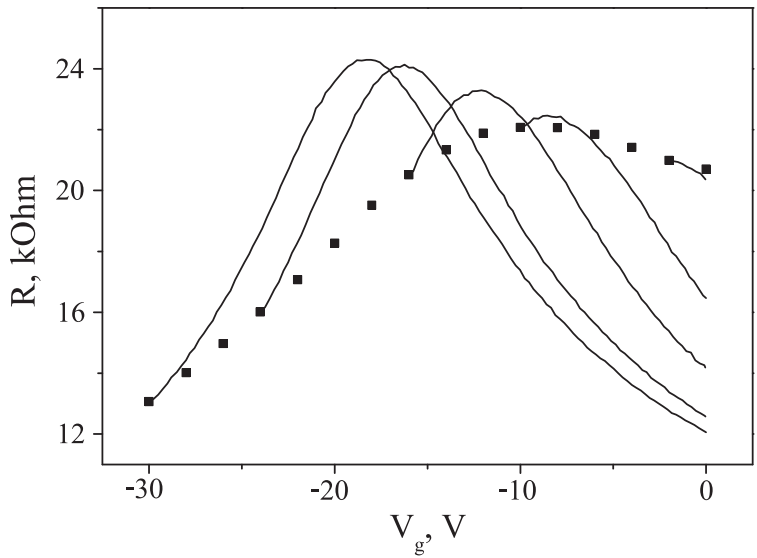}
\caption {Resistance of a sample without PMMA processing after one hour waiting at different gate voltages in the presence of toluene (points). The curves are fast sweeps to $V_{g}=0$ (sweep rate $\sim 1$\,V/s) from each point. They show that the position of the resistance peak (Dirac point) depends on the initial value of $V_g$.}
\label{fig:three}
\end{figure}

To explore this result in more detail we looked at the time dependence of the doping process. As can be seen in the inset to \ref{fig:four}, all curves share the same rate of change. This suggests that the rate-limiting step in this process is not the diffusion of toluene (or other molecules) on the graphene surface, which would depend on the strength and polarity of the applied voltage. The curves can be fitted empirically by: 
\[
R(t)=R(0)+A_{1}\exp{\left(-\frac{t}{\tau_{1}}\right)}+A_{2}\exp{\left(-\frac{t}{\tau_{2}}\right)}\, ,
\]
where $A_1$, $A_2$, $\tau_1$ and $\tau_2$ are constants. There are two characteristic times: a short time $\tau_1\simeq 200$\,s and a long time $\tau_2\simeq 4400$\,s. The shorter time $\tau_1$ puts a lower limit on the sweep rate of $V_g$. Therefore, in order to observe both timescales associated with the doping we change the gate voltage over a time much smaller than $\tau_1$. \ref{fig:three} shows the value of the resistance (points) of the sample after waiting for one hour at a particular $V_g$. From each point, $V_g$ has been rapidly swept to $V_g=0$ (shown as curves in the figure). The sweeping time from $V_g=-30$\,V to 0 is less than $30$\,s ($\ll\tau_1$), so the system is not able to relax back to its equilibrium state. The curves in the figure, therefore, show the position of the Dirac point for each particular initial value of $V_g$. From this it can be seen that the Dirac point shifts as the initial value is changed.

The slow characteristic time $\tau_2$ suggests that a chemical reaction is the mechanism of transferring electrons to graphene. Furthermore, the dependence of the Dirac point position on gate voltage suggests that this reaction is influenced by electric field and is therefore electrochemical in nature \cite{Pinto10030624}. When a toluene molecule loses one electron to graphene it becomes oxidised to a radical. The toluene radical is highly reactive and as such can take part in many different chemical reactions with other species that are present in the system, particularly water. Water is very likely to be present \cite{JoshiJPCM22} as it can strongly bond to the silica surface and is not readily removed by annealing \cite{LangeJCS20}. It is not possible to determine which particular reaction takes place (and it is likely that there are several of them occurring), however, we can experimentally determine if the reaction is electrochemical in nature by measuring the associated redox energy level $\varepsilon_R$. This level will depend on the experimental conditions, the concentration of toluene and products of its reaction, so we must compare it with the energy difference of $\sim$1.1\,eV (3.9\,eV) between the HOMO (LUMO) level in toluene and the Fermi level in graphene, $\sim$4.6\,eV \cite{YuNL9}, which would be the energy gap seen if only molecular doping occurred \cite{Pinto10030624}.

\begin{figure}
\includegraphics[width=.6\textwidth]{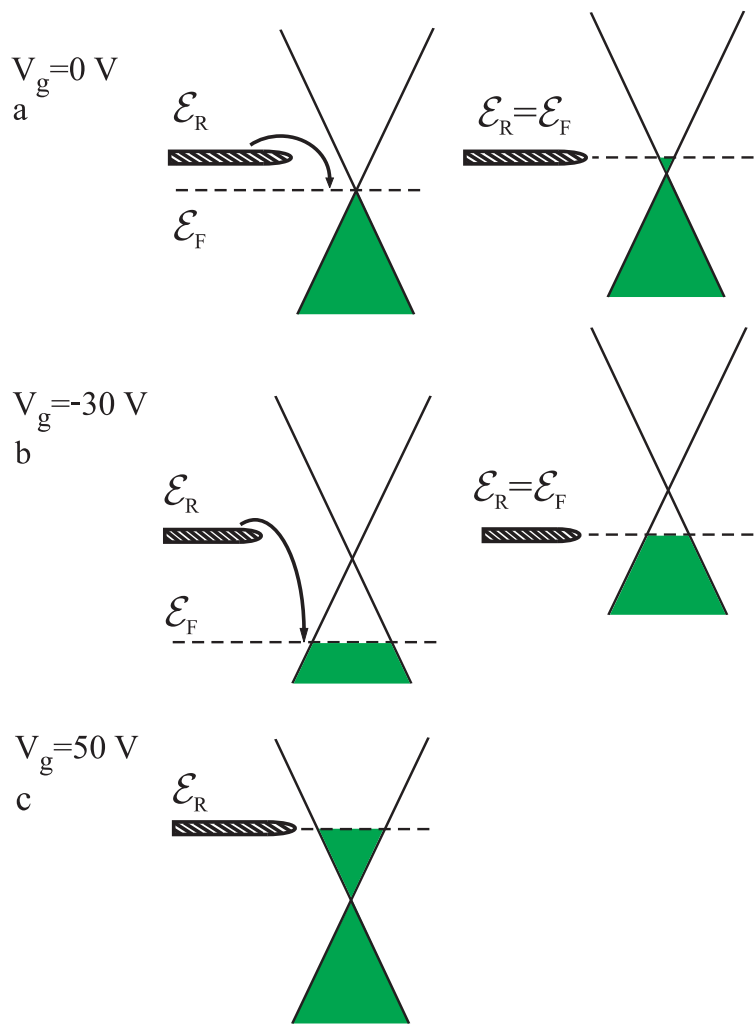}
\caption {Energy diagram of graphene bands during electrochemical doping. The Fermi level $\varepsilon_{F}$ and redox level $\varepsilon_{R}$ are shown before (left) and after (right) doping by toluene. Three regimes are shown: (a) $\varepsilon_{R}>\varepsilon_{F}$, $V_{g}=0$; (b) $\varepsilon_{R}\gg\varepsilon_{F}$, negative $V_g$; (c) $\varepsilon_{R}<\varepsilon_{F}$, positive $V_g$.}
\label{fig:six}
\end{figure}

\ref{fig:six} shows the mechanism of electrochemical doping of graphene. If $\varepsilon_R>\varepsilon_F$ then doping of graphene will occur until the condition $\varepsilon_R=\varepsilon_F$ is met at equilibrium. The larger $\varepsilon_R$ is compared to $\varepsilon_{F}$, the more electrons will be transferred to the graphene and the larger the shift in $V_g$ will be. (We do not expect an increase in the transfer rate, as, although the density of states in graphene increases linearly away from the Dirac point, the transfer rate is dominated by the energy barrier for the reaction.) This is seen in \ref{fig:four} as a larger resistance change at negative applied gate voltage. This mechanism also explains the dependence of Dirac point position on gate voltage shown in \ref{fig:three} as it is defined by the initial values of $\varepsilon_R$ and $\varepsilon_F$, i.e. the initial number of transferred electrons. However, if $\varepsilon_R\leq\varepsilon_F$ then no doping can occur and hence there is no dependence of the Dirac point position on $V_g$.

\begin{figure}
\includegraphics[width=.6\textwidth]{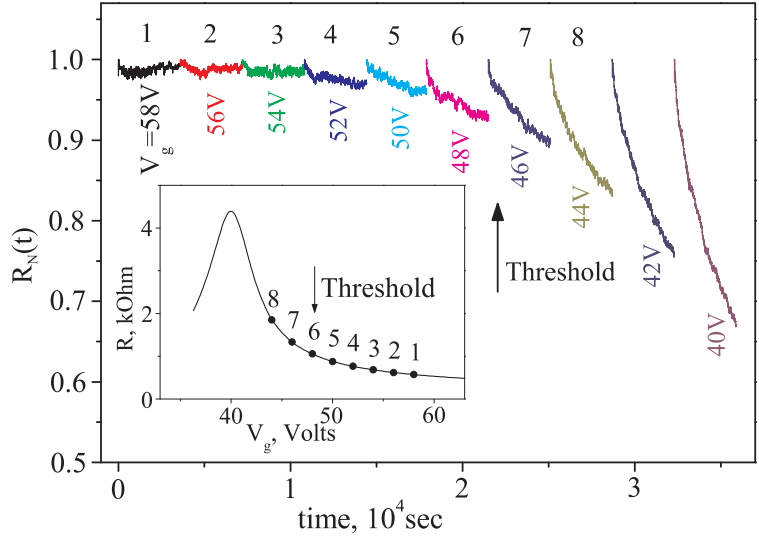}
\caption{Resistance as a function of time as the gate voltage is changed from an initial value of $+60$\,V. Between each rapid 2\,V change in $V_g$ the system is left for one hour. The inset shows the corresponding positions as a function of $V_g$ in the case where the gate voltage is swept very slowly compared to $\tau_2$.}
\label{fig:five}
\end{figure}

In order to establish the presence and magnitude of the redox level we investigated the dependence of the Dirac point position on $V_g$. First, in an inert atmosphere $V_g$ is set to a large positive value of $+60$\,V to ensure that the condition $\varepsilon_{R}<\varepsilon_{F}$ will be satisfied when toluene is introduced. \ref{fig:five} shows the time dependence of the normalised resistance in the presence of toluene vapour. The gate voltage is fixed for one hour between being rapidly changed by 2\,V decrements towards zero. For gate voltages down to $\sim 48$\,V little change is observed. In contrast, below $48$\,V an exponential change is observed. We ascribe this onset of exponential change to the alignment of the Fermi level in the graphene with the redox level. This threshold gate voltage corresponds to an energy of $\sim 0.1$\,eV (as it is 10\,V with respect to the Dirac point), which is an order of magnitude smaller than the energy threshold of 1.1\,eV expected for molecular doping. This result, along with the slow characteristic time $\tau_2$ points strongly to an electrochemical origin. (The origin of the faster time $\tau_1$, however, remains unclear.)

The electrochemical nature of the doping by dipolar molecules explains why H$_2$O but not naphthalene would cause a gate-voltage dependent Dirac point. Naphthalene does not participate in electrochemical reactions under our experimental conditions (requiring significantly higher temperatures), and therefore the energy associated with its doping is the HOMO level gap of $\sim$1\,eV to the Fermi level in graphene which lies well beyond the range of accessible gate voltages. In contrast, water vapour readily undergoes this type of reaction and as such the threshold energy for doping can be significantly reduced, in a similar way to toluene.


In summary, we have demonstrated that the doping of graphene by toluene can be understood in terms of an electrochemical reaction mechanism. We have shown that toluene acts as a donor, but that the transfer of electrons can be controlled by an electric field. This was demonstrated by a hysteretic dependence of the resistance of a graphene transistor as a function of the applied gate voltage in the presence of toluene vapour. We have also shown that the dipolar nature of the molecule is a factor, the same effect being observed for another dipolar molecule, water, but not for the nonpolar molecule naphthalene. By measuring the point of onset of the doping we were able to determine the magnitude of the redox energy level to be $\sim 0.1$\,eV for our experimental conditions, an energy much smaller than that expected from the simple doping mechanisms considered earlier.

\begin{acknowledgement}
The authors thank H.~Pinto, R.~Jones, A.~S.~Shytov, S.~J.~Green and D.~W.~Boukhvalov for useful discussions, P. R. Wilkins for technical support, and the EPSRC (grant numbers EP/G036101/1 and EP/G041482/1) for funding.
\end{acknowledgement}

\bibliography{tolbib}

\end{document}